\documentstyle[aps,twocolumn]{revtex}

\begin{document}
\twocolumn[\hsize\textwidth\columnwidth\hsize
\csname@twocolumnfalse%
\endcsname

\draft
\title{Behavior of tumors under nonstationary theraphy}
\author{O. Sotolongo-Costa$^{(a,d)}$, L. Morales Molina $^{(b)}$\thanks{Corresponding author.email: lmolina@reduc.cmw.edu.cu \hspace{1cm} fax: 5332287020}, D. Rodr\'{i}guez
Perez$^{(c)}$, J. C. Antoranz$^{(c,d)}$, M. Chac\'{o}n Reyes.$^{(b)}$}
\address{$^{(a,d)}$ Department of Theoretical Physics, Havana University, 10400, Havana, Cuba
\\
$^{(b)}$ Department of Physics, University of
Camaguey, 74650, Camaguey, Cuba
\\
$^{(c)}$ Department of mathematical Physics, UNED, Madrid,
\\
$ ^{(d)}$ Henri Poincare Chair of Complex Systems, Havana University}

\maketitle
\date{\today{}}

\begin{abstract}
\ We present a model for the interaction dynamics of lymphocytes-tumor cells
population. This model reproduces all known states for the tumor. Futherly,
we develop it taking into account periodical immunotheraphy treatment with
cytokines alone. A detailed analysis for the evolution of tumor cells as a
function of frecuency and theraphy burden applied for the periodical
treatment is carried out. Certain threshold values for the frecuency and
applied doses are derived from this analysis. So it seems possible to
control and reduce the growth of the tumor. Also, constant values for
cytokines doses seems to be a succesful treatment.
\end{abstract}

\pacs{PACS: 87.22.-q  
            02.30.Hq  
            05.45.Xt  
\\ 
\ Keywords: Immunotherapy- Tumor- Cytokine- Modelling- Ordinary differential equation- Coupled Oscillators}
\vskip 1cm
]
\section{Introduction}

\ Cancer is one of the leading research areas, since this desease is a main
cause of death. Surgery and chemotherapy are unsuccessful in many cases.
Today the principal efforts are addressed to search new treatment strategies
e.g. in immunotherapy (see Refs.\cite{Forys,Paneta} and references therein).
In this case we refer to the use of cytokynes to stimulate the immune
system. This is a protein hormone produced by activated lymphocytes which
mediate both natural and specific immunity. The use of cytokines alone to
boost the immune system represent one of the methods more commonly used in
immunotherapy. The temporal evolution for this treatment comprises different
steps: by supplying a starting dose of cytokines the rate of lymphocytes
begins to increase due to the immunological reaction \cite{Forys} reaching a
maximum value; afterwards the lymphocytes begin to decay because of the
decrease of cytokines concentration inside the body until it reaches normal
values. In the following we shall call this time the activation period of
the immune system. This proccess repeats again between any two succesive
dose suplies. We consider these injections separated in time by the dosage
period. Otherwise, it would provoke an overdose or failure in the treatment
for shorter and longer time, respectively. \newline
\ This work is devoted to understand the temporal tumor behavior when a
periodical immunotherapy treatment is provided. Besides, we want to explore
the set of parameter values to reproduce with our model the same features
presented in \cite{Paneta}(like short tem oscillation in tumor size and long
term tumor relapse). \newline
\ For this reason we reformulate the predator-prey model including new terms
in the model which give account of tumor agressiveness, the diffusion of
lymphocytes and the effect caused by cytokines on the tumor (Section 2). We
analyse the model from the mathematical and biological points of view
(Sections 3 and 4). We study in detail tumor evolution for different
treatment regimes (Section 5) and, in the last section, we discuss our
results and suggest some options for improving the model.

\section{The model}

An extensive review of all models of tumor-immune system dynamics \cite
{Paneta,Bellomo,Bell,Chaplain1,Chaplain2,Kuznetsov1,Kuznetsov2,Kuznetsov3,Delisi,Lefever1,gonzalez} as far as
we know was done. We agree with the idea that such dynamics is determined by
a competence of interacting species resembling the predator-prey model. The
main difference between our model and the clasical deterministic model (e.g.
Bell 1973)\cite{Bell} is the inclusion of new terms taking into account, (1)
the death of lymphocytes due to the increase of malignant cells population,
(2) the flux of lymphocytes towards the place of local interaction and (3)
the effect produced by the application of cytokine doses. \newline
\ Let $X $ and $Y $ denote respectively the number of malignant and
lymphocyte cells. The rate of malignant cells $\left( \frac{dX}{dt}\right) $
is given by: 
\begin{equation}  \label{1}
\frac{dX}{dt}=\, \, aX-\, \, bXY
\end{equation}
We assume a growth rate proportional to $X $ and a decrease rate
proportional to the frecuency of interaction with lymphocytes. Coefficients
are $a $ and $b $, respectively, where $a $ is tissue dependent.

\ On the other hand, the growth rate of lymphocytes $\left( \frac{dY}{dt}%
\right) $ is described by: 
\begin{equation}  \label{2}
\frac{dY}{dt}=\, \, dXY-\, \, fY-\, \, kX+u
\end{equation}
\newline
It is proportional to the interaction with malignant cells and also to the
flux per unit time of lymphocytes to the place of interaction. These effects
are represented by the first and fourth terms in the right-hand side of
equation \ref{2}. This last term characterizes the difussion process of
lymphocytes that takes place in the surroundings of the tumor assuming a
constant lymphocytes flux \cite{Kuznetsov1}. On the other hand, the decrease
rate depends on two factors: natural death and growth of malignant cells
related to the effective area of tumor interacting directly with the
lymphocytes. These are given by the second and third terms of this same
equation where $f $ and $k $ are their respective coefficients of
proportionality.

\ In order to introduce the effects produced by the treatment with cytokines
in the proccess of activation of the immune system, we add a periodical
function that mimics the periodical dosage. As a first approximation, we
propose the function \ $F\cos ^{2}{\omega t} $, where $\omega $ is the
frecuency of the periodical behavior for cytokines inside the body. \ The
modified Eq \ref{2} will be given by the expression

\begin{equation}  \label{3}
\frac{dY}{dt}=\, \, dXY-\, \, fY-\, \, kX+u+F\cos ^{2}{\omega t}
\end{equation}

Taking into account Eq.\ref{1}, we get the following system of differential
equations.

\begin{eqnarray}
\frac{dX}{dt} & = & \, \, aX-\, \, bXY  \label{4} \\
\frac{dY}{dt} & = & \, \, dXY-\, \, fY+Q  \label{5} \\
\, \, Q & = & -\, \, KX+u+Fcos^{2}(\omega t)  \label{6}
\end{eqnarray}

\ where $Q$ is a function that picking up all the news contributions
respecting the standard predator-prey model. \newline
with $X(0)=X_{0} $ and $Y(0)=Y_{0} $ as initial conditions.

\ From the analysis for $Q=0 $ the system of Eq.\ref{4}, Eq.\ref{5}and Eq.%
\ref{6} reduces to the predator-prey model, which has as equilibrium points
a saddle point at the phase portrait origin and a center in the first
quadrant of the phase diagram\cite{Holden}. This center reflects the
oscillating behavior of the competence between both predator and prey, which
is characterized by an oscillation frecuency (number of cycles per unit time
around the center of the phase diagram). \ Taking the reverse of this
frecuency as a characteristic time $t_{0}=\frac{1}{\sqrt{af}} $ and
rescaling the equations of the system of Eq.\ref{4}, Eq.\ref{5}and Eq.\ref{6}
by means of the following scaling parameters:

\begin{eqnarray}
\, \, t & = & \, \, t_{0}\tau  \nonumber \\
X & = & X^{\prime}x  \nonumber \\
Y & = & Y^{\prime}y  \nonumber 
\end{eqnarray}

\ the system given by the Eq.\ref{4}, Eq.\ref{5}and Eq.\ref{6} become:

\begin{eqnarray}
\, \, \frac{X^{\prime}}{t_{0}}\dot{x} & = & \, \, axX^{\prime}-\, \,
bxyX^{\prime}Y^{\prime}  \label{7} \\
\frac{Y^{\prime}}{t_{0}}\dot{y} & = & \, \, (dxX^{\prime}-\, \,
f)yY^{\prime}+Q  \label{8} \\
\, \, Q & = & \, \, F\cos ^{2}(\omega \, \, t_{0}\tau)+u-K\, \, X^{\prime}x
\label{9}
\end{eqnarray}

\ sustituting the scaling parameters by the following values: 
\begin{eqnarray}
\, \, t_{0} & = & \frac{1}{\sqrt{af}}  \nonumber \\
X^{\prime}& = & \frac{\sqrt{af}}{d}  \nonumber \\
Y^{\prime}& = & \frac{\sqrt{af}}{b}  \nonumber 
\end{eqnarray}

\ we get the following rescaled equations

\begin{eqnarray}
\frac{dx}{d\tau } & = & \alpha \, \, x-\, \, xy\, \,  \label{10} \\
\frac{dy}{d\tau } & = & \, \, xy-\frac{1}{\alpha }\, \, y-\, \,
kx+\sigma+V\cos ^{2}(\beta \, \, \tau )  \label{11}
\end{eqnarray}

\ with $x(0)=x_{0} $ and $y(0)=y_{0} $ as initial conditions. \newline

\ where $V=\frac{Fb}{af} $, \ $k=\frac{K}{bd}\sqrt{af} $,

\ $\alpha=\sqrt{\frac{a}{f}}$, $\sigma=\frac{ub}{af} $   and 
$\beta =\frac{\omega }{\sqrt{af}} $\newline
\ Analysing in detail the prior expressions we can interpret these
parameters as follows: \newline

\ From $V$ it is deduced that its value depends on the net value for doses ($%
F$) and on the action of the immune system on malignant cells ($b$). Hence
it would represent an effective value of all doses employed in the
activation of lymphocytes, namely, those necesary for activating the
lymphocytes that take action directly against the tumor cells. \newline
\ On the one hand, $k$ is directly proportional to $K$ which accounts for
the negative effects exerted on the population of lymphocytes due to the
size of the tumor and it is inversely proportional to the recognition ($d$)
and attack ($b$) frequencies of the immune systen to malignant cells. So, we
infer that the inverse value of this parameter ($1/k$) gives, {}``in some
sense{}'', the control exerted by the immune system over the aggressiveness
of the tumor due to its size. \newline
\ From the similarity of the expressions for $\sigma $ and $V$, taking into
account the remarks done for $V$ we interpret that $\sigma $ which depends
on the action of the immune system on malignant cells ($b$) represent of all
lymphocytes flux $u$ those effective value involved in the attack on tumor
cells.\ Finally, $\alpha $ and $\beta $ are directly related to the
proliferation of malignant cells and the frecuency of treatment,
respectively.

\section{Dynamics and Stability Analysis without treatment}

\ The system of Eq.\ref{10} and Eq.\ref{11} corresponding to $V=0 $( no
treatment) is an autonomous system given by the following equations:

\begin{eqnarray}
\frac{dx}{d\tau } & = & \alpha \, \, x-\, \, xy\, \,  \label{12} \\
\frac{dy}{d\tau } & = & \, \, xy-\frac{1}{\alpha }\, \, y-\, \, kx+\sigma
\label{13}
\end{eqnarray}
\ with $x(0)=x_{0} $ and $y(0)=y_{0} $ as initial conditions. \newline
\ Susbstituting the Eq.\ref{12} in the Eq.\ref{13} we get the differential
equation

\begin{equation}  \label{14}
\frac{d^{2}x}{d\tau ^{2}}+(\frac{1}{\alpha }-x-\frac{1}{x}\frac{dx}{d\tau })%
\frac{dx}{d\tau }=(k-\alpha )x^{2}+\sigma x
\end{equation}

\ with $x(0)=x_{0}$, $\dot{x}(0)=v_{0}$ as initial conditions.

This equation is similar to that describing the motion of a particle in a
force field \cite{gonzalez}, whose potential is:

\begin{equation}
\ U(x)=-\frac13(k-\alpha)x^3-\frac12\sigma x^2  \label{15}
\end{equation}
\newline
This potential has two extremes given by \newline
\centerline{\( x_{1}=0\hspace {1cm}\mbox {and}\hspace {1cm}x_{2}=\frac{\, \,
\sigma-1}{\, \, k-\alpha } \)} These extreme points depend on $\alpha $, $k $
and $\sigma $ as:

\centerline{\( \, \, \sigma>1\rightarrow \left\{ \begin{array}{c}
\, \, x_{1}=0\hspace {1cm}\mbox {minimum}\vspace {0.5cm}\\
\hspace {0.5cm}\frac{\, \, k}{\alpha }>1,\hspace {1cm}x_{2}>0\hspace {1cm}\mbox {maximum}\vspace {0.5cm}\\
\hspace {0.5cm}\frac{\, \, k}{\alpha }<1,\hspace {1cm}x_{2}<0\hspace {1cm}\mbox {maximum}
\end{array}\right.  \)} \ and for \newline
\centerline{\( \, \, \sigma<1\rightarrow \left\{ \begin{array}{c}
\, \, x_{1}=0\hspace {1cm}\mbox {maximum}\vspace {0.5cm}\\
\hspace {0.5cm}\frac{\, \, k}{\alpha }>1,\hspace {1cm}x_{2}>0\hspace {1cm}\mbox {minimum}\vspace {0.5cm}\\
\hspace {0.5cm}\frac{\, \, k}{\alpha }<1,\hspace {1cm}x_{2}<0\hspace {1cm}\mbox {minimum}
\end{array}\right.  \)}\newline
We only consider motion for $x>0$, the suitable potential fields describing
the motion of the particle are depicted in Fig.\ 1a and Fig.\ 1b. In Fig.\
1a , the maximum represents an unstable point for the particle motion,
contrary in Fig.\ 1b the particle oscillates around the minimum. \newline
\ The analysis of fixed points in the phase space for Eq.\ref{12} and Eq.\ref
{13} shows two steady-states. A fixed point is $L_{0}=(0,\alpha \sigma )$
with associated eigenvalues 
\begin{equation}
\lambda _{\pm }=\frac{\alpha ^{2}(1-\sigma )-1}{2\alpha }\pm \left| \frac{%
\alpha ^{2}(1-\sigma )+1}{2\alpha }\right|  \label{16}
\end{equation}
\ For $\sigma <1$, $L_{0}$ is a saddle point while for $\sigma >1$ is a
stable node.\newline
\ The other fixed point is $L_{1}=\left( \frac{1-\sigma }{\alpha -k},\alpha
\right) $ with associated eigenvalues given by:

\begin{equation}  \label{17}
\, \, \lambda _{\pm }=\frac{k-\alpha \sigma}{2\alpha (\alpha -k)}\pm \sqrt{%
\left[ \frac{k-\alpha \, \, \sigma}{2\alpha (\alpha -k)}\right]
^{2}-(1-\sigma)}
\end{equation}

\ The real part of this eigenvalue is zero for $\frac{k}{\alpha }=\sigma $
with $v_{0}<1 $. \newline
\ When condition $(\alpha -k)^{2}-\frac{k}{\alpha }>-1 $ is fulfilled, we
get two values for $\sigma_{c} $ solutions of \newline
\newline
\centerline{\( \alpha ^{2}\sigma_{c}^{2}-2\alpha [k-2\alpha (\alpha
-k)^{2}]\sigma_{c}+k^{2}-4\alpha ^{2}(\alpha -k)^{2}=0 \)}\newline
\ given by

\begin{eqnarray}
\, \, \sigma_{c}=\frac{k}{\alpha }-2\alpha ^{2}\left( 1-\frac{k}{\alpha }%
\right) ^{2}\hspace{2cm}\nonumber
\\
\pm 2\alpha \arrowvert 1-\frac{k}{\alpha }\arrowvert \sqrt{%
\alpha ^{2}\left( 1-\frac{k}{\alpha }\right) ^{2}-\frac{k}{\alpha }+1}\label{18}
\end{eqnarray}
\newline
\ defining the region of complex eigenvalues and focus-like behavior.

\ The analysis of these eigenvalues provides a rich dynamics. For the case $%
\frac{k}{\alpha }<1 $ we have different situations. The states and its stability for the second fixed point are depicted schematically in
Fig.\ 2. \newline
If $\sigma<\frac{k}{\alpha } $ ($Re\lambda _{\pm }>0 $), we have an unstable
focus or node depending on the parameter value $\sigma_{c} $ relative to
those given by Eq.\ref{18}. \newline
On the contrary, stable behavior (focus or node) appears when $\frac{k}{%
\alpha }<\sigma<1 $ ($Re\lambda _{\pm }<0 $). \newline
Now, if $1<\sigma $ the fixed point corresponds to a negative population of
malignant cells, with no physical meaning. \newline
For the case $\frac{k}{\alpha }>1 $ and $\sigma<1<\frac{k}{\alpha } $ ($%
Re\lambda _{\pm }<0 $) the critical point moves to the second quadrant of
the phase diagram, being discarded as before. \newline
For values of $\sigma $ in the ranges $1<\sigma<\frac{k}{\alpha } $ and $1<%
\frac{k}{\alpha }<\sigma $ we get a saddle point ($\lambda _{+}>0 $ and $%
\lambda _{-}<0 $), whose separatrix splits the phase portrait into stable
and unstable zones as can be seen in Fig.5. \ In all the cases the dynamics
in the phase diagram is represented by a homeomorfism\cite{Holmes} between
two fixed points.

\section{Biological Significance}

\ So far, we have presented a detailed analitical study of the linear
stability of our model when $V$ is set equal to zero. The interpretation of
this preliminary results will give us the esential features of the system.%
\newline
Let us start with the case $\frac{k}{\alpha }<1$. \newline
\ For $\sigma <\frac{k}{\alpha }$ the system evolves towards a state of
uncontrolable tumor growth(see Fig.\ 3a) This case can be interpreted as a 
{\em recurrence} like behavior \cite{Michelson,Wheldom} very similar to
q-switching oscillations observed in physical phenomena as, for example, in
lasers. On the contrary, when $\frac{k}{\alpha }<\sigma <1$, our
system evolves towards a controlable mass of malignant cells in a damped
oscillating way (Fig.\ 3b). This state is considered by some authors as a
dormant state\cite{Kuznetsov1,Kuznetsov2,Kuznetsov3,Michelson,Wheldom}. 
\newline
\ However in both cases, there exist populations of malignant cells that
grow towards a state in which immunological activity has been suppressed. In
the first case this happens for any initial conditions, while in the second
it only happens for an initially weak inmunological response (Fig.\ 4a and
4b). Let us now analyse the reverse condition $\frac{k}{\alpha }>1$. \newline
In this case there are two possible ranges for $\sigma $: \newline
\centerline{\( 1<\frac{k}{\alpha }<\sigma \) and \( 1<\sigma<\frac{k}{\alpha
} \)}\newline
In both situations we are in the presence of a saddle point which means that
for populations of cancer cells below the horizontal separatrix the dynamics
is irreversible: this curve represents the critical amount of malignant
cells for a fixed population of lymphocytes. \newline
\ This situation is similar to the case analised before for a weak immnune
system as an initial condition: the population of malignant cells grows
towards a value such that the immunological response is reduced to zero
(Fig.\ 5). This would represent a state where illness is not the cause of
death but leaves the body unprotected against other diseases. \newline
\ However for malignant cells above the horizontal separatrix it is possible
to observe regression of tumor as has been reported in clinical
experiments(see Ref.\cite{Michelson} and references therein)

\section{Stability analysis with treatment and biological implications}

\ The system represented by Eq.\ref{10} and Eq.\ref{11} with $V\neq 0$ (cytokines
doses amplitude) can be analised as an autonomous system\cite{Hao}. The
procedure consists in substituting the oscillating function $\cos {\beta t}$
of the driven term $F\cos ^{2}{\beta t}$ in the Eq.\ref{11} by a new
variable $u$, which is a solution of the second order differential equation
of a linear oscillator 
\begin{equation}
\frac{d^{2}u}{dt^{2}}+\beta ^{2}u=0  \label{19}
\end{equation}
(where $u(0)=1$, $\dot{u}(0)=0$) which can be written as two linear coupled
differential equations 
\begin{eqnarray}
\frac{dz}{d\tau } &=&-\beta ^{2}u  \label{20} \\
\frac{du}{d\tau } &=&z  \label{21}
\end{eqnarray}
(with $u(0)=1$, $z(0)=0$)\newline
\ Then Eq.\ref{10} and Eq.\ref{11} become: 
\begin{eqnarray}
\frac{dx}{d\tau } &=&\alpha \,\,x-\,\,xy\,\,  \label{22} \\
\frac{dy}{d\tau } &=&\,\,xy-\frac{1}{\alpha }\,\,y-\,\,kx+\sigma +Vu^{2}
\label{23} \\
\frac{dz}{d\tau } &=&-\beta ^{2}u  \label{24} \\
\frac{du}{d\tau } &=&z  \label{25}
\end{eqnarray}
\ with $x(0)=x_{0}$, $y(0)=y_{0}$, $u(0)=1$, $z(0)=0$ as initial conditions. 
\newline
This system presents the critical points $L_{0}^{\ast }=(0,\alpha \sigma
,0,0)$ and $L_{1}^{\ast }=\left( \frac{1-\sigma }{\alpha -k},\alpha
,0,0\right) $ whose projection in the $y$-$x$ plane coincides with those
critical points of the unperturbed system (Eq.\ref{12} and Eq.\ref{13}) with
the same eigenvalues given by the Eq.\ref{16} and Eq.\ref{17} plus the new
conjugate pair $\lambda _{\pm }=\pm \,\,i\beta $. \newline
In this case we are in presence of a center manifold where solutions can be
expanding or contracting, i.e., the asymptotic stability analysis carried
out before loses its validity, needing more complex developments.\newline
\ In order to avoid such complexity and gain a better comprehension, we may
consider this system like a couple of one linear (Eq.\ref{19}) and one
nonlinear (Eq.\ref{14}) oscillators, allowing us a more intuitive
interpretation of the different regimes. Thus the changes from recurrent to
dormant states of tumor cells in the periodical dosage regime can be
interpreted as a lock of the unstable oscillations of the nonlinear
oscillator imposed by the linear one. We can understand the complex behavior
of coupled oscillators by representing its dynamics as a function of control
parameters $V$ and $\beta $\cite{Hao,Strogatz}. In order to depict it we plot,
in the parameter space ($V\mbox {vs}\beta $), those points for which tumor
growth is uncontrolable. \newline
\ In the case $\frac{k}{\alpha }<1$ for $\sigma <\frac{k}{\alpha }$, for
effective value of doses and frecuencies higher than certain threshold, the
system can revert from uncontrolable growth to a treatment controled
population. Namely, for every set of parameter values $\alpha $,$k$ and $%
\sigma $, there are threshold values for $\beta $ and $V$ which split the
parameter space (see Fig\ 6) into two zones corresponding to uncontrolable
and controlable growth of malignant cells. The Fig\ 6 was generated for a
fixed set of initial conditions. Although the behavior of the parameter
space for different initial conditions is qualitatively the same, the
threshold values show strongh dependence of the initial conditions.
\\
\
From this result and taking into account the meaning of the parameters $
\beta $ and $V$, we can infer that treatment is specific for each patient
and kind of tumor since threshold doses values depend on the immunological
response of each individual, on the malignant cells population at the
begining of the treatment and also on the rate of proliferation of the
tissue. Besides, the existence of threshold values reflects the fact that
reaching controlable populations of malignant cells is only possible by
mantaining a minimal dose above certain threshold given by the continuous
line depicted in Fig\ 6, which can be well fitted by a hiperbolic function. 
\newline
For effective doses and frecuencies below these threshold values, the system
behaves qualitatively the same as without treatment. However, contrary to
this statement it is also found that for low frecuencies the growth of
malignant cells can be controlled in spite of being below the threshold
values (see, for details, Fig\ 6). \newline
There exists a \char`\"{}paradoxical\char`\"{} phenomenon observed in
experiment and the clinic, consisting in the fact that the enhacement of the
inmune system with immunotherapy stimulates tumor growth\cite{Outzen}, which
could be explained, {}``in some way{}'', by this result, i.e., why, for
fixed doses burden , the growth becomes uncontrolable at given frecuencies
above those localized in the region of controlable growth of malignant cells
(Fig.\ 6). Such values would represent an optimal treatment as it reduces
doses burden and treatment frecuency. \newline
Also, for higher frecuencies with small $V$, growth of cancer cells can be
controlled as shown in Fig\ 6. The observance of these optimal values would
be important because of the negative effects produced when cytokines
concentration reaches above a critical concentration\cite{Rosenstein,Lissoni}%
.. \newline
\ Now, increasing the amplitud of the effective dose value for a fixed
frecuency above the threshold, a malignant cells population reduction is
obtained, in spite of an uncontrolable growth being observed for some higher
doses burden (see Fig\ 6). \newline
\ On the other hand, setting $V$ to some value and varying the frecuency
from zero to higher values, different behaviors are reproduced. At zero
treatment frecuency, tumor cells population is controlable with the lowest
values of doses burden (Fig.\ 6). However, for frecuencies different from
zero, we find zones of recurrent and dormant growth of tumor cells. The
population of cancer cells controled under treatment presents an oscillating
behavior (see Fig.\ 7)\cite{Paneta}. \newline
For values of the parameters satisfying $\frac{k}{\alpha }<\sigma <1$ (that
we interpret as a dormant state), malignant cells population can be reduced
by increasing effective doses. On the other hand, varying the frecuency for
fixed effective doses values, an oscillating behavior for the population of
malignant cells is obtained, as in the previous case. \newline
In all these cases, regrowth of malignant cells takes place after treatment
interruption \cite{Kuznetsov3}. This can be easily understood if we take
into account that the population of malignant cells with zero value ($x=0$)
represents, in the mechanical analogue (Eq.\ref{14}), an unstable point (a
potential maximum, as that shown in Fig.\ 1b). This means that any variation
will lead the system towards a minimal potential position. Therefore for a
residual population slightly greater than zero, a regrowth of tumor cells
will take place after the treatment.\newline
Analysing the behavior when $\frac{k}{\alpha }>1$ we arrive at the following
results. The range of values with physical sense for $\sigma $, i.e. $1<%
\frac{k}{\alpha }<\sigma $ and $1<\sigma <\frac{k}{\alpha }$ allows only two
critical points: a stable node and a saddle point. In this case dynamics in
the phase portrait is the same as that without treatment. There are no
possible changes in the dynamics, so treatment is useless. \newline
Hence, for $\frac{k}{\alpha }>1$, only initial conditions determine the
final outcome of tumor evolution, irrespective of the applied treatment.

\section{Conclusions}

\ In this work we intend to give a new focus to the dynamics of tumor growth
in a periodical regime of immunotherapy. We explain such dynamics
considering this system as two coupled oscillators, namely, the competence
between the immune system and malignant cells analized as a nonlinear
oscillator coupled with a linear one that simulates the treatment.\newline
\ This simple model allowed us to describe all possible states in wich a
tumour can be found. It also presented some of the features found in tumor
dynamics, outlined by some authors, such as the existence of short term
oscillations of tumor size as well as the long term tumor relapse. On the
other hand, this model gives the dependence of tumor growth on some
parameter values related to the treatment: the frecuency and amount of
applied doses. We conclude from this study that the evolution of tumor
submitted to immunotherapy has a strong dependence on these parameter
values. \newline
In some cases, growth of malignant cells can be reverted with inmunotherapy
treatment. Corresponding threshold values are obtained for treatment
frequency and dose above which growth is stopped and malignant cells
population reduced.\newline
Also, for those inmunological parameters for which a stable population of
malignant cells exists, the size of the dormant tumor can be reduced by
increasing dose burden.\newline
It was shown, as well, that for certain relation among the parameter values,
tumor presents a recurrent behavior with or without treatment.\newline
In all these cases, when a reduction of the tumor is possible, best results
are obtained for low constant values of the dose. Nevertheless in all cases
previously analyzed, after the interruption of the treatment, tumor regrowth
is observed. \newline
This would confirm the fact that treatment with immunotherapy using
cytokines alone is not succesful enough in the treatment against cancer\cite
{Paneta}. Therefore another kind of therapy would be required. \newline
\ As a way to improve the model, we propose the introduction of other terms
taking into account effects produced by stochastic perturbations due to
enviromental conditions\cite{Chaplain2,Kuznetsov1,Lefever1}. Some authors atribuit to
these perturbations the main cause of possible jumps from stable to unstable
behavior in tumor growth dynamics. This will be considered in a future work.
\

\begin{figure}[tbp]
\caption{Potential barriers in which the particle moves. (a) $\protect\alpha %
=1$, $k=1.5$ and $\sigma=3$. (b) $\protect\alpha =2$, $k=0.2$ and $\sigma=0.25$
..}
\label{fig1}
\end{figure}

\begin{figure}[tbp]
\caption{States and its stability for the second fixed point as a function of parameters $\protect%
\alpha $, $k$, $\sigma$.}
\label{fig2}
\end{figure}

\begin{figure}[tbp]
\caption{Evolution of malignant cells on time without treatment. (a) $%
\protect\alpha =2$, $k=0.2$ and $\sigma=0.05$ with $x_{0}=2.1$ and $y_{0}=2.7$%
. (b) $\protect\alpha =2$, $k=0.2$ and $\sigma=0.25$ with $x_{0}=5.3$ and $%
y_{0}=6.7$.}
\label{fig4}
\end{figure}

\begin{figure}[tbp]
\caption{Phase portraits (lymphocytes population versus malignant cells).
(a) $\protect\alpha =2$, $k=0.2$ and $\sigma=0.09$. (b) $\protect\alpha =2$, $%
k=0.2$ and $\sigma=0.25$.}
\label{fig5}
\end{figure}

\begin{figure}[tbp]
\caption{Saddle point in the phase portrait for values of parameters $%
\protect\alpha =1$, $k=1.5$, $\sigma=3$.}
\label{fig3}
\end{figure}

\begin{figure}[tbp]
\caption{Growth behavior of malignant cells with treatment for $\protect%
\alpha =2$, $k=0.2$ and $\sigma=0.05$ with $x_{0}=5.3$ and $%
y_{0}=6.7$ depicted in the parameter space ($V$, $%
\protect\beta $). Uncontrolable growth (gray points). Controlable growth
(white points). Black solid line (hiperbolic function $%
V=0.10478+0.00044/(0.05343+\protect\beta )^{2.7313}$).}
\label{fig6}
\end{figure}

\begin{figure}[tbp]
\caption{Limit cycles for $V=0.25$ and different values of parameter $%
\protect\beta $ in the phase portrait.}
\label{fig7}
\end{figure}

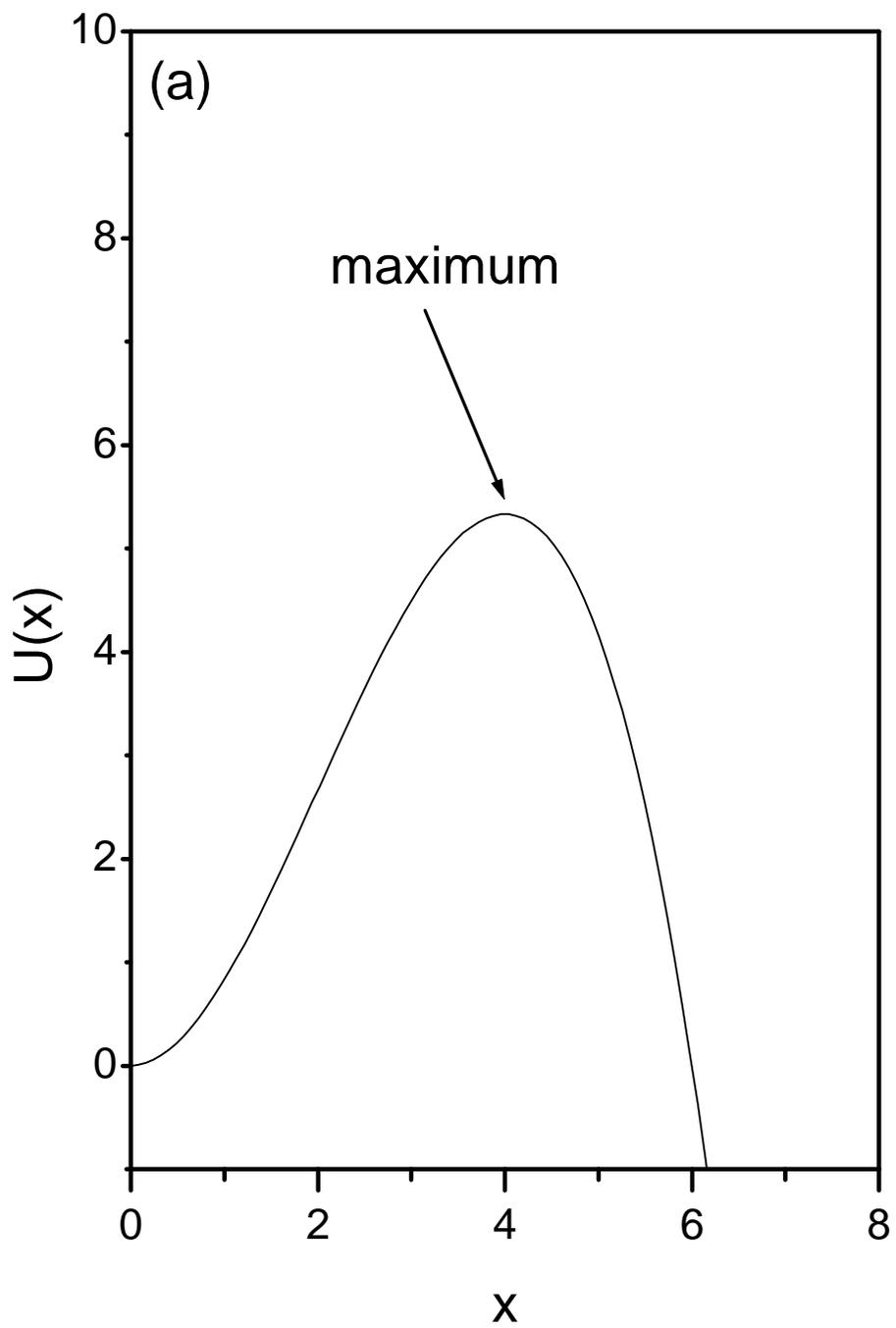
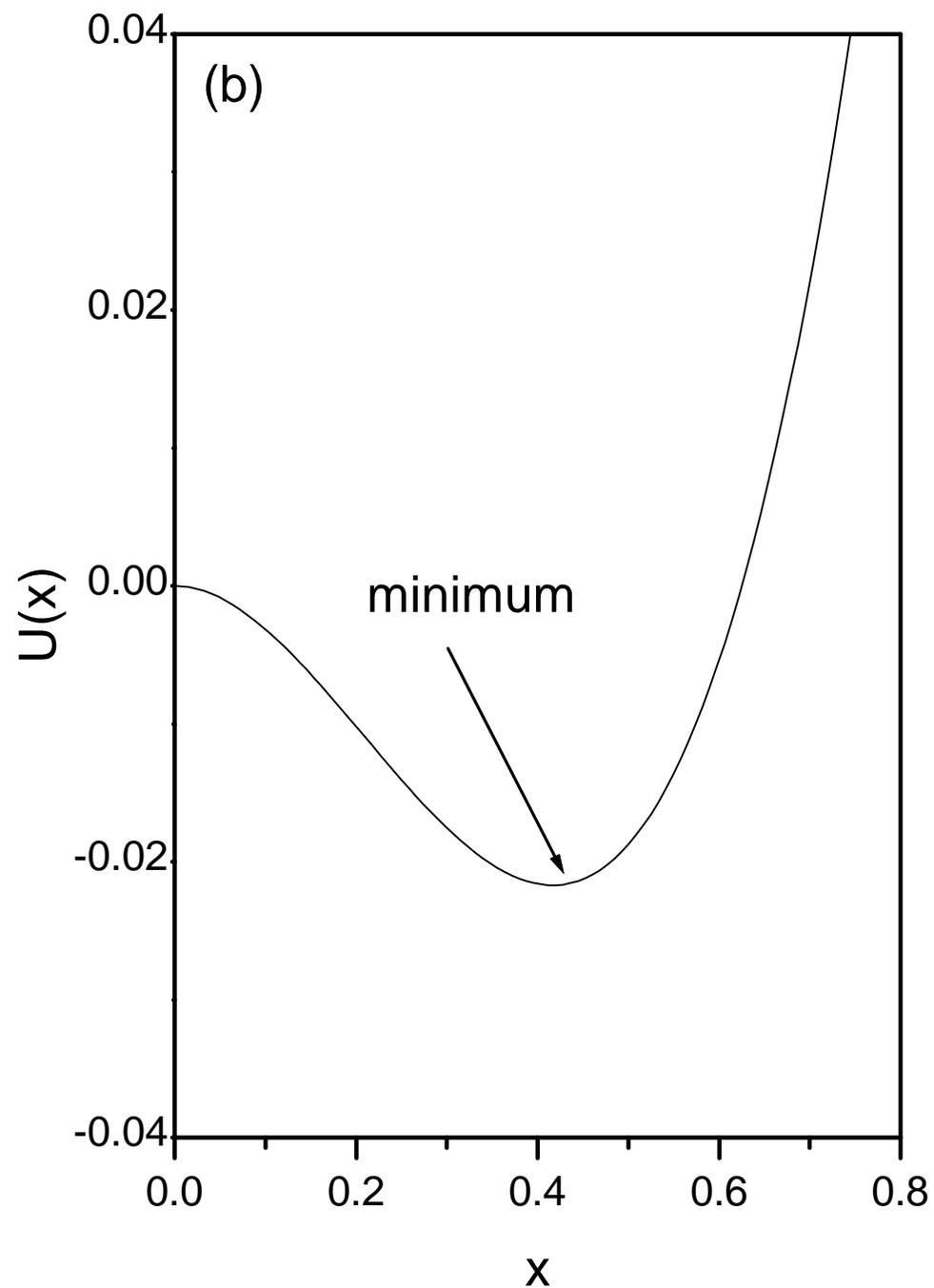
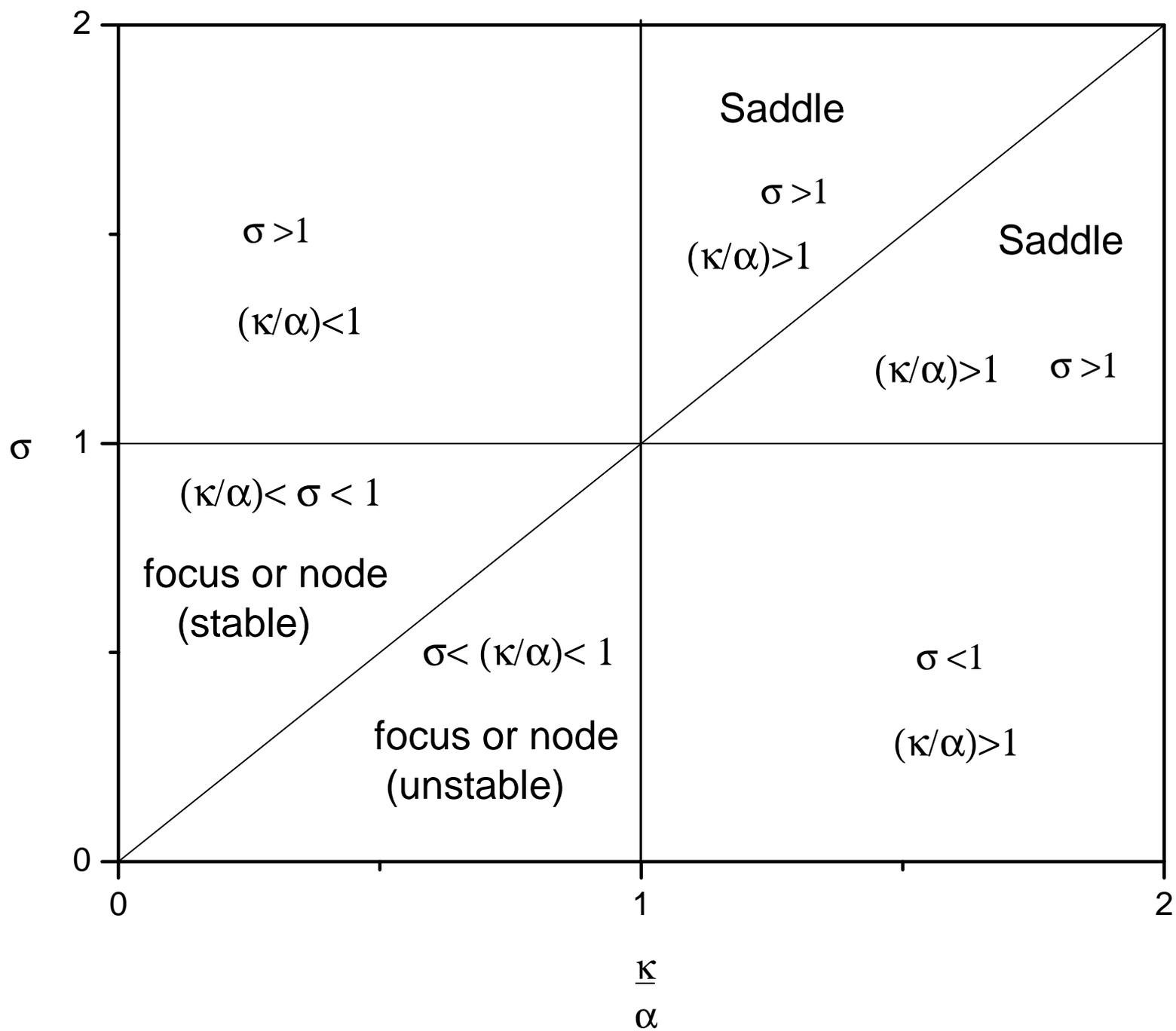
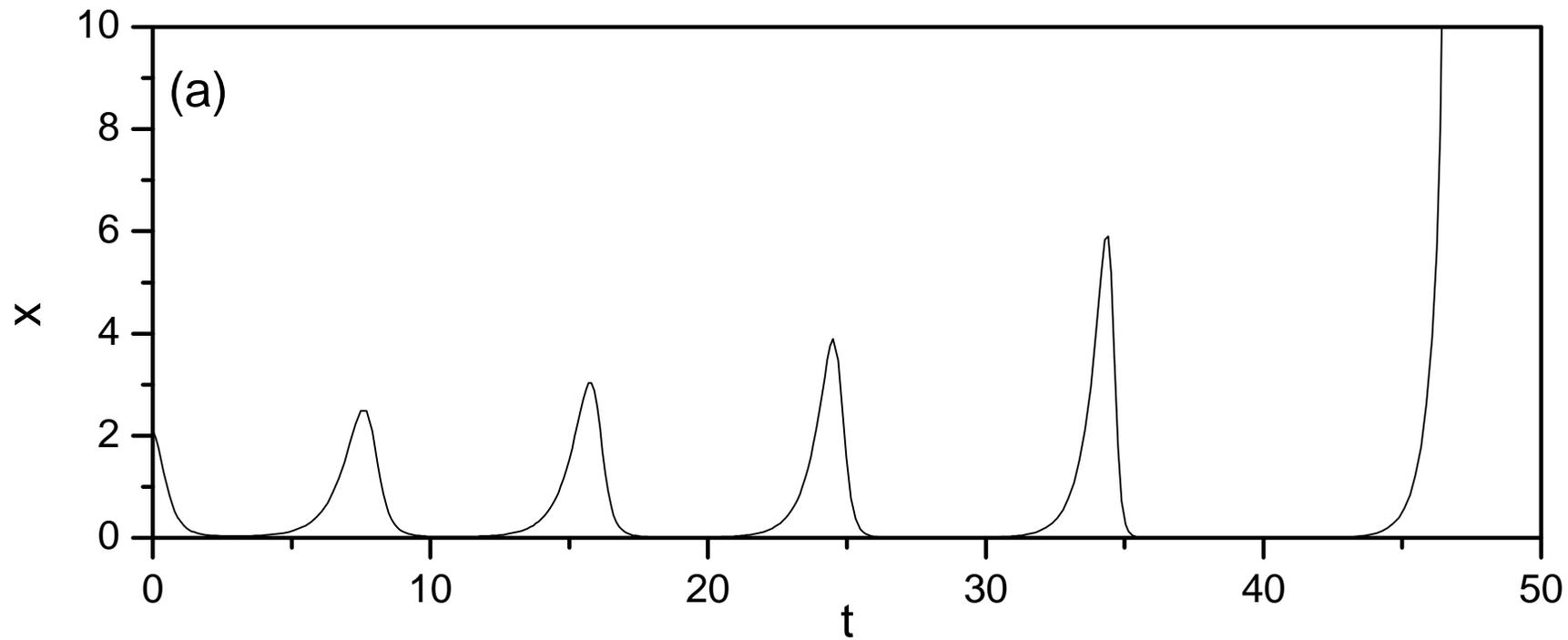
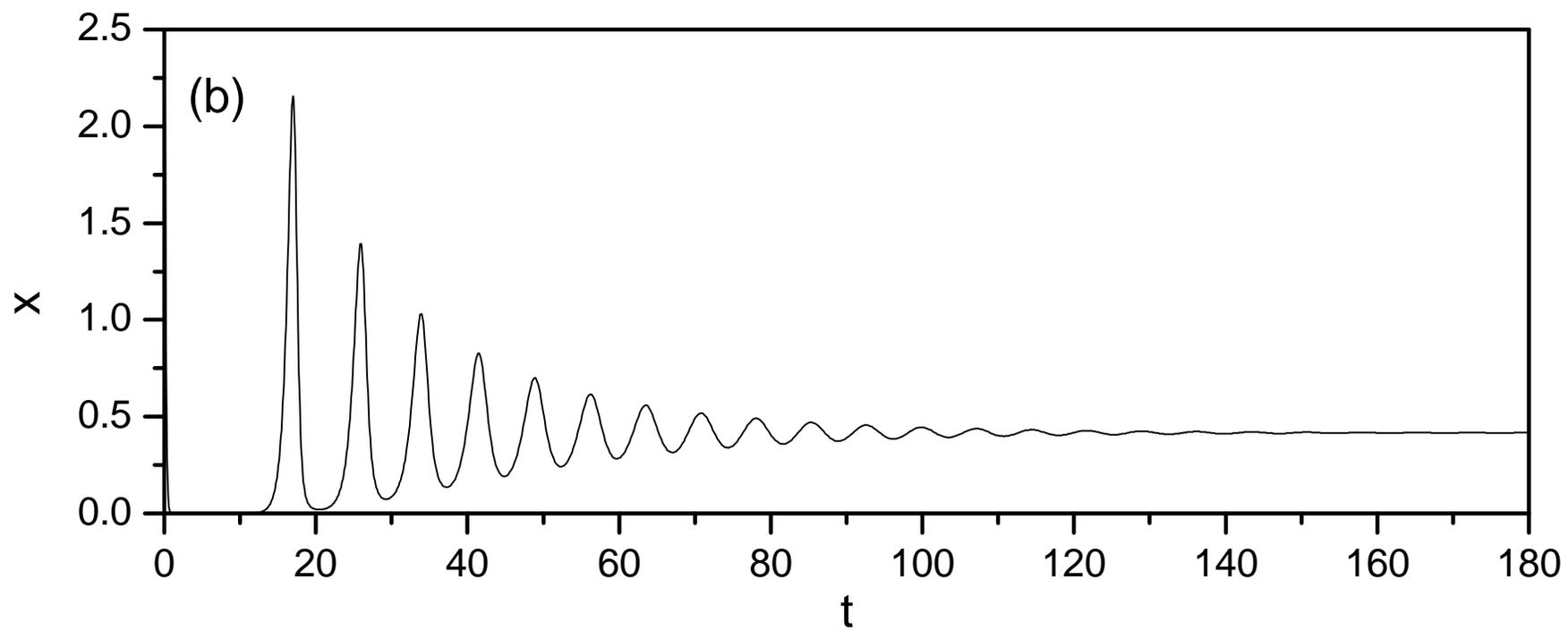
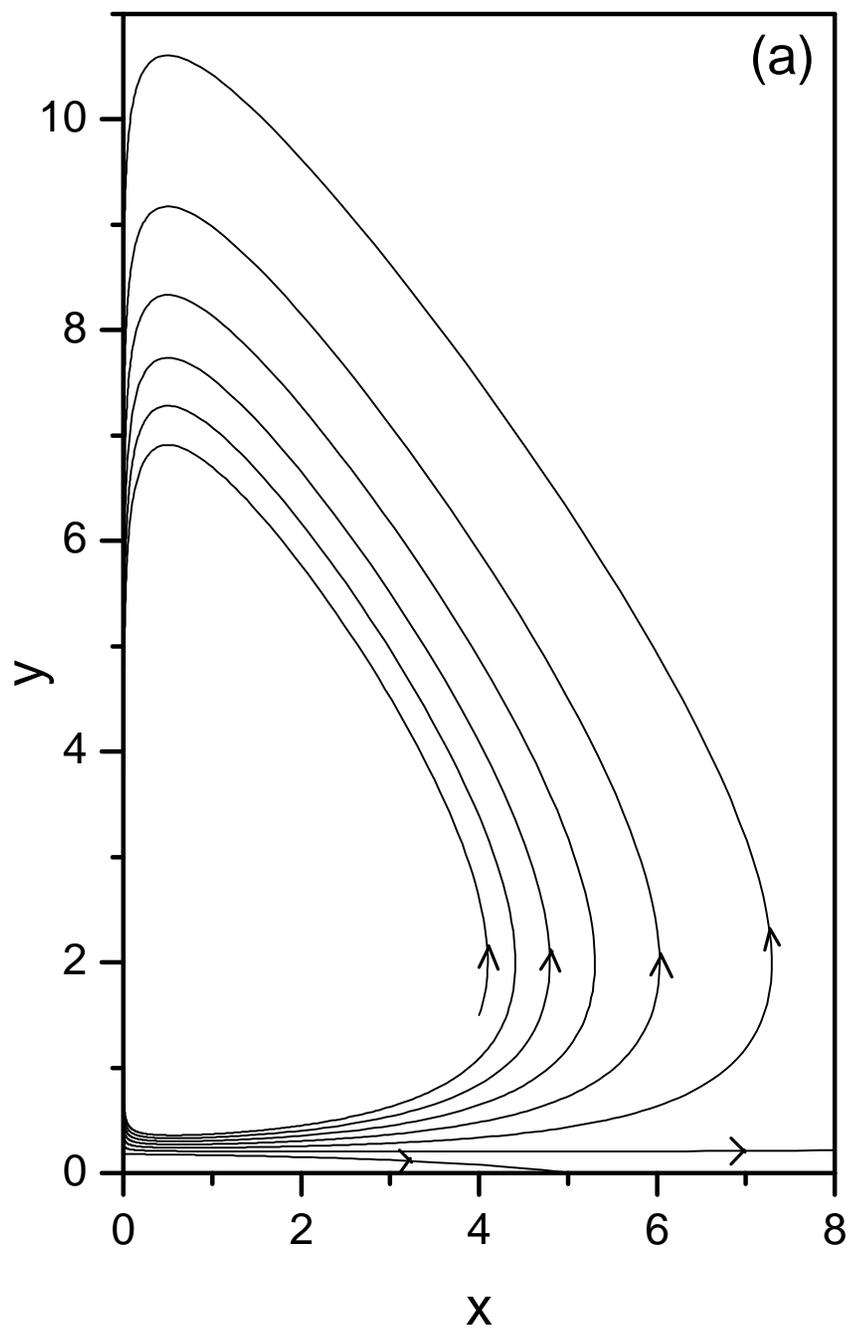
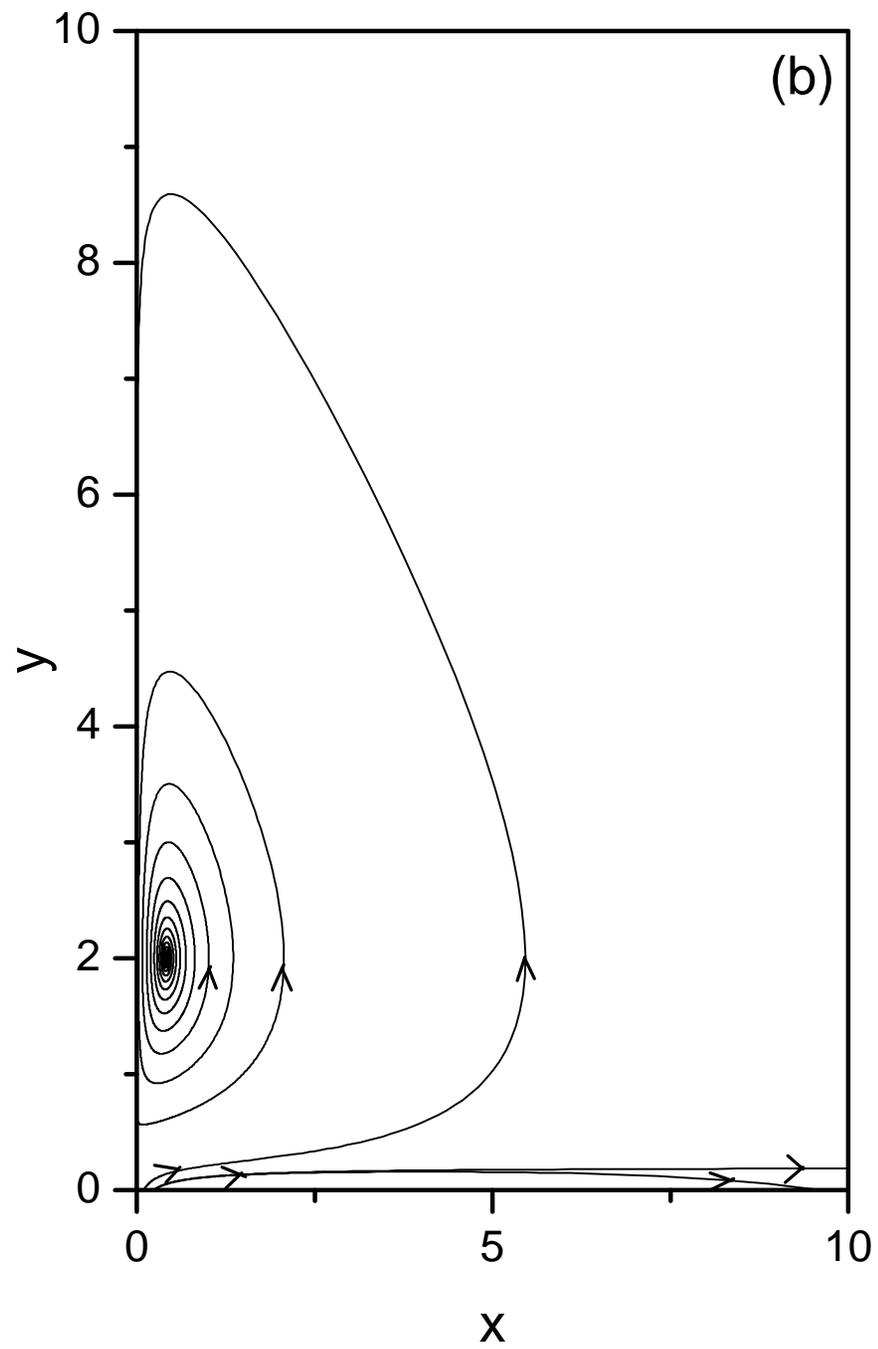
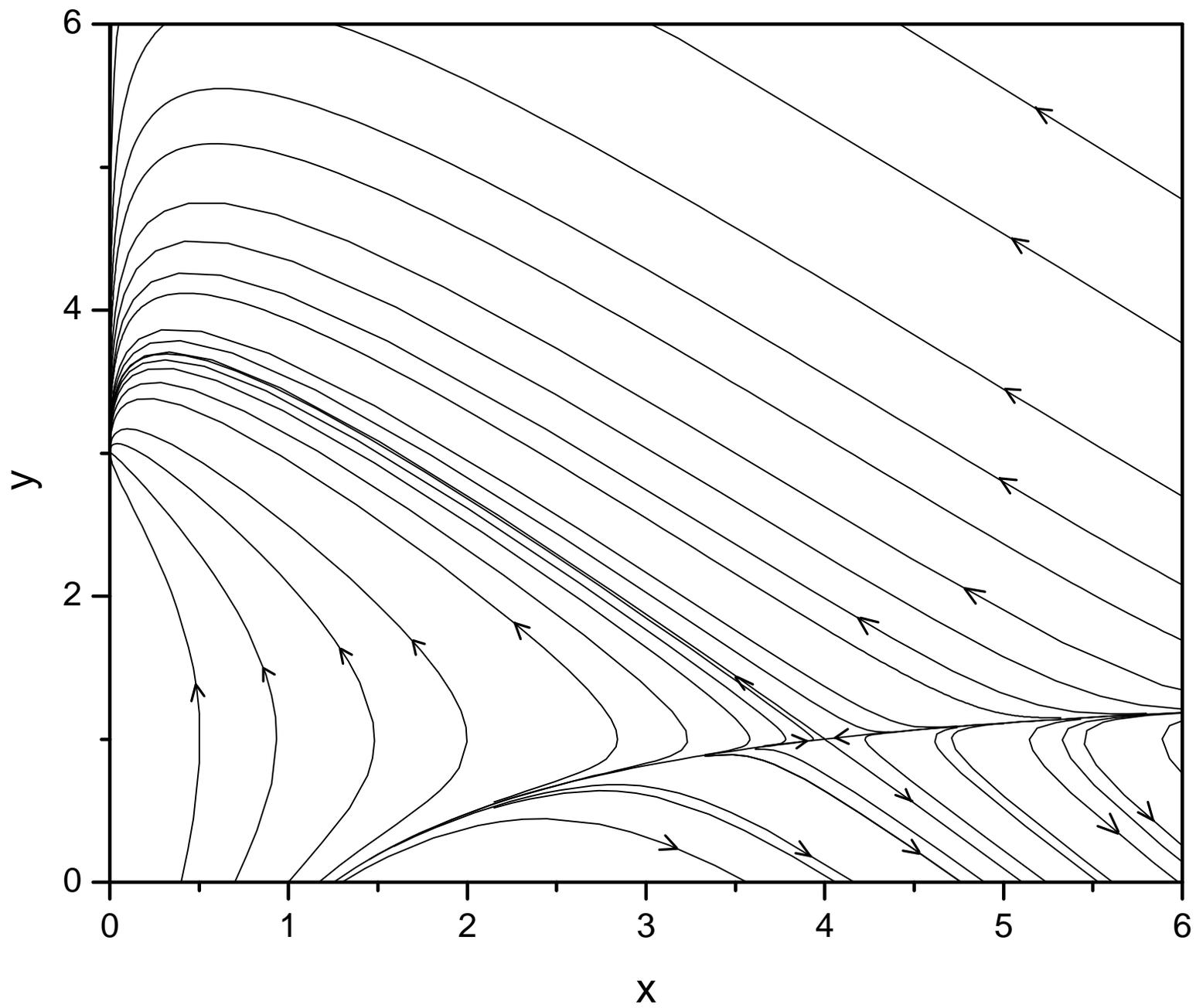
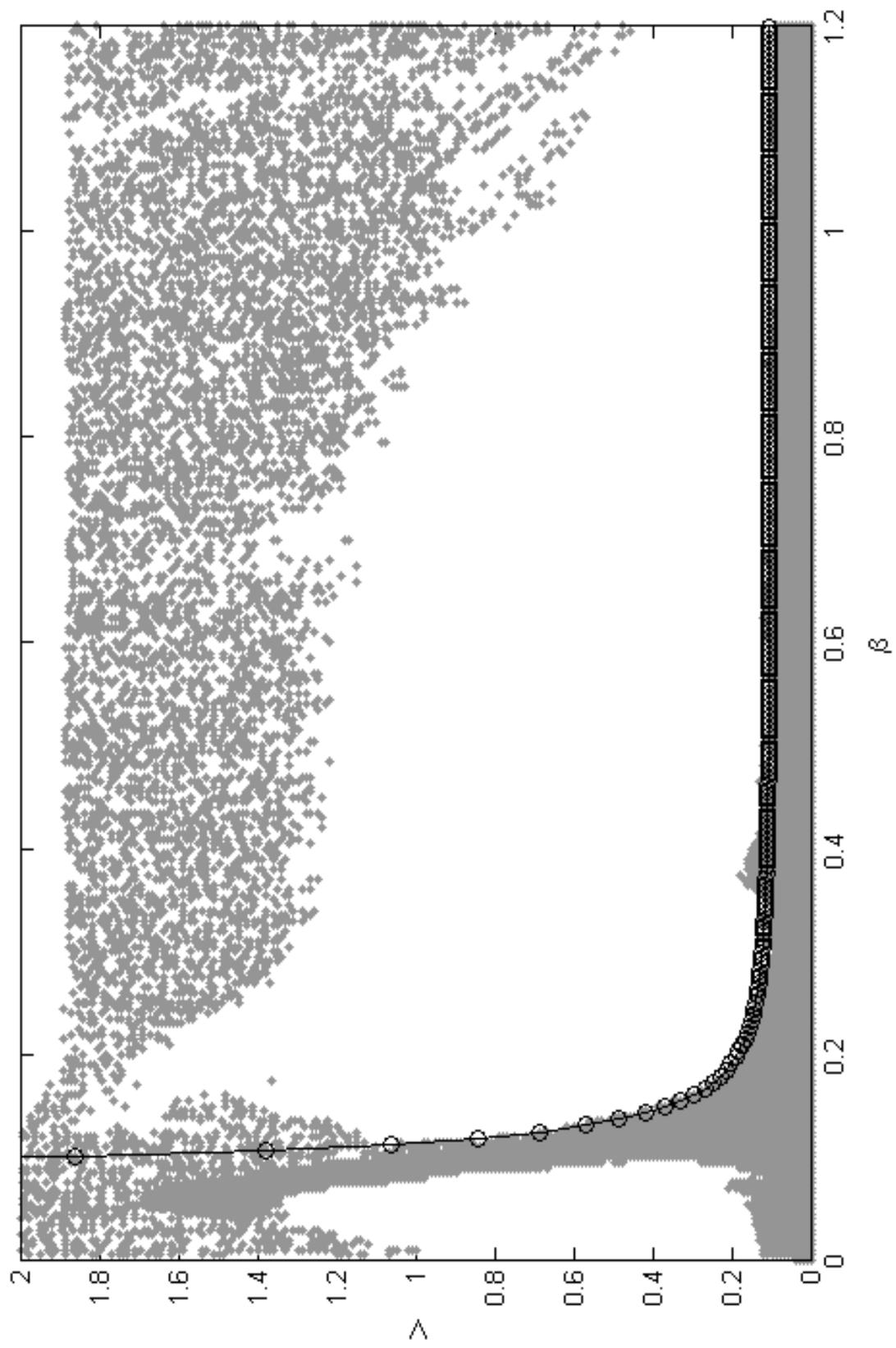
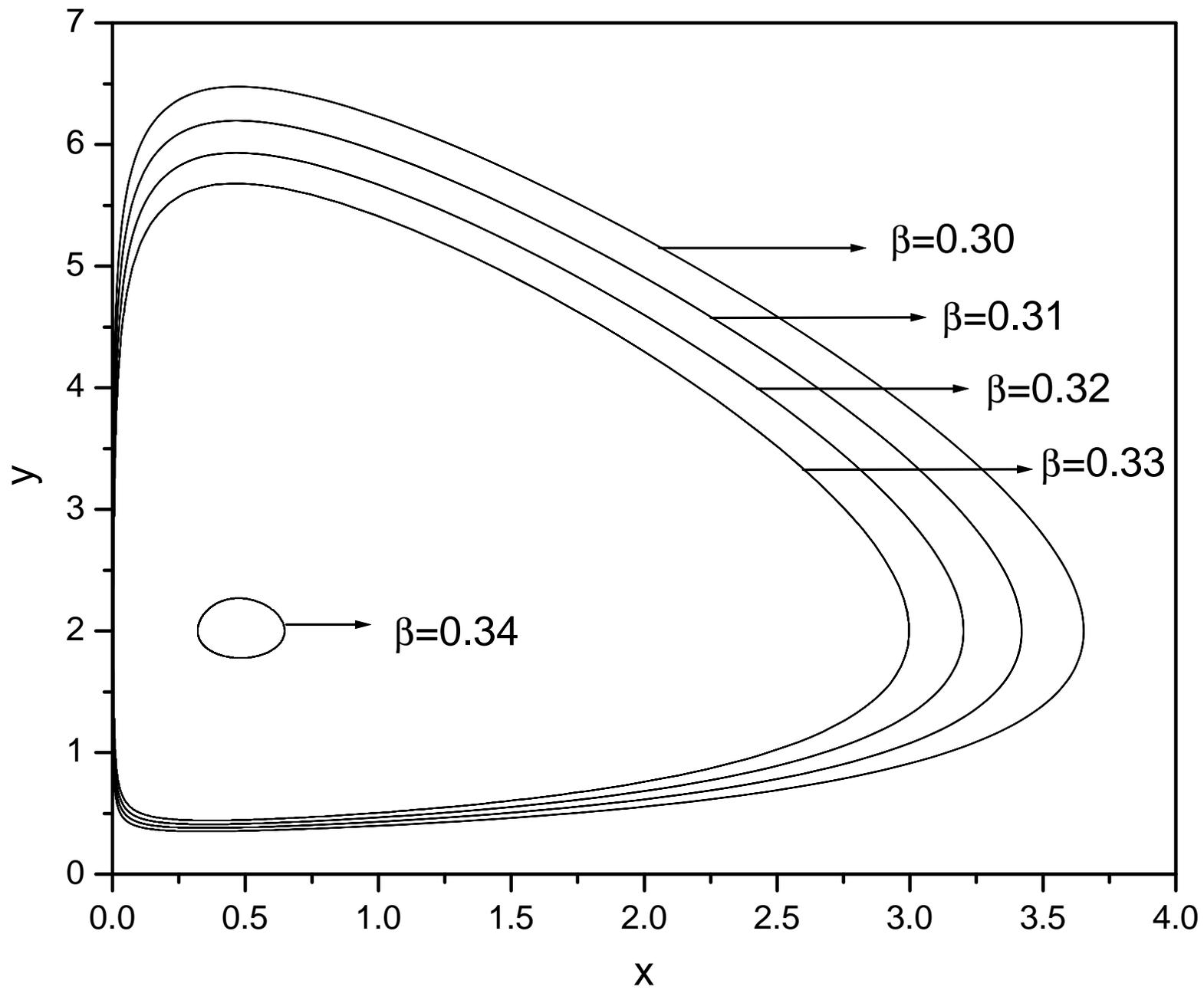
\end{document}